%% file: RewStrat.tex
\documentclass[copyright,creativecommons]{eptcs}
\providecommand{\event}{SR 2013} 
\usepackage{breakurl}             

\usepackage{verbatim}

\usepackage{amsmath}
\usepackage{mathrsfs}
\usepackage{amssymb}
\usepackage{url}

\input{minmacros}

\title{A rewriting point of view on strategies} 

\author{H\'el\`ene Kirchner\\
Inria\\
Domaine de Voluceau - Rocquencourt\\
BP 105 - 78153 Le Chesnay Cedex France\\
\url{e-mail: Helene.Kirchner@inria.fr}\\
}

\def\titlerunning{A rewriting point of view on strategies}
\def\authorrunning{H\'el\`ene Kirchner}

\begin{document}
\maketitle

\begin{abstract}
This paper is an expository contribution reporting on published work.
It focuses on an approach followed in the rewriting community
to formalize the concept of strategy.
Based on rewriting concepts, 
several definitions of strategy are reviewed  and connected:
in order to catch the higher-order nature of strategies,
a strategy is defined as a proof term expressed in the rewriting logic or
in the rewriting calculus;
to address in a coherent way  deduction and
computation,  a strategy is seen as a subset of 
derivations; 
and to recover the definition of strategy  in sequential path-building
games or in functional programs, 
a strategy is considered as a partial function that associates to a 
reduction-in-progress, the possible next steps in the reduction sequence.

\end{abstract}

\section{Introduction}

Strategies frequently occur  in automated deduction and reasoning
systems and more generally are used to express complex designs for 
control in modeling, proof search, program transformation, SAT solving
or security policies. 
In these domains, deterministic rule-based computations or deductions are often not sufficient to
capture complex computations or proof developments. A formal mechanism is needed, for
instance, to sequentialize the search for different solutions, to check context
conditions, to request user input to instantiate variables, to process subgoals in a
particular order, etc. This is the place where the notion of strategy comes in.

This paper deliberately focuses on an approach
followed in the rewriting community to formalize a notion of
strategy relying on
rewriting logic~\cite{marti-oliet00}  and
rewriting calculus~\cite{rhoCalIGLP-I+II-2001} 
that are powerful formalisms to
express and study uniformly  computations and deductions in automated
deduction and reasoning systems.
Briefly speaking, rules describe local transformations and strategies describe the control of  rule application.
Most often, it is useful to distinguish between rules for computations, where
a unique normal form is required and where the strategy is fixed, and
rules for deductions, in which case no confluence nor termination is required 
but an application strategy is necessary.
Regarding rewriting as a relation and considering abstract rewrite
systems leads to consider derivation tree exploration:
derivations are computations and 
strategies  describe selected computations. 

Based on rewriting concepts, that are briefly recalled in
Section~\ref{rewriting},
several definitions of strategy are reviewed  and connected.
In order to catch the higher-order nature of strategies,
a strategy is first defined as a proof term expressed in rewriting logic in Section~\ref{RewritingLogic}
then in rewriting calculus in Section~\ref{RewritingCalculus}.  In Section~\ref{ARS},
a strategy is seen as a set of paths in a derivation tree;
then to recover the definition of strategy  in sequential path-building
games or in functional programs, 
a strategy is considered as a partial function that associates to a reduction-in-progress, the possible next 
steps in the reduction sequence. 
In this paper, the goal is to show the progression of ideas and definitions
of the concept, as well as their correlations.

\section{Rewriting}
\label{rewriting}

Since the 80s, many aspects of rewriting have been studied in
automated deduction, programming languages, equational theory decidability,
program or proof transformation, but also in various domains such as
chemical or biological computing, plant growth modeling, etc.
In all these applications, rewriting definitions have the same basic
ingredients.
Rewriting transforms syntactic structures that may be
words, terms, propositions, dags, graphs, geometric objects like
segments, and in general any kind of structured objects.
Transformations are expressed with  patterns or rules. Rules are built
on the same syntax but with an additional set of variables, say ${\cal  X}$, 
and with a binder $\Rightarrow$, relating the  left-hand side and the right-hand
side of the rule, and optionally with a condition or constraint that
restricts the set of values allowed for the variables.
Performing the transformation of a syntactic structure $t$ is applying
the rule labeled $\ell$ on $t$, which is basically done in three steps:
(1) match to select a redex of $t$ at position $p$ denoted $t_{|p}$ (possibly modulo some axioms, constraints,...);
(2) instantiate the rule variables by the result(s) of the matching
substitution $\sigma$;
(3) replace the redex by the instantiated right-hand side. Formally:
$t$ rewrites to $t'$ using the rule $\ell : l \rewrite r$ if
$t_{|p} = \sigma(l)$ and $t' = t[\sigma(r)]_p$.
This is denoted $t \rew{}{}{p,\ell,\sigma} t'$.

In this process, there are many possible choices: the rule itself,
the position(s) in the structure, the matching substitution(s).
For instance, one may choose to apply a rule concurrently at all 
disjoint positions where it matches, or using matching modulo an
equational theory like associativity-commutativity, or also
according to some probability.

\section{Rewriting logic}
\label{RewritingLogic}

The Rewriting Logic is due to J.~Meseguer and N.~Mart{\'\i}-Oliet~\cite{marti-oliet00}.\\
As claimed on \url{http://wrla2012.lcc.uma.es/}:

{\em Rewriting logic (RL) is a natural model of computation and
  an expressive semantic framework for concurrency, parallelism,
  communication, and interaction. It can be used for specifying a wide
  range of systems and languages in various application fields. It
  also has good properties as a metalogical framework for
    representing logics. In recent years, several languages based on
  RL (ASF+SDF, CafeOBJ, ELAN, Maude) have been designed and
  implemented. }

In Rewriting Logic, 
the syntax is based on a set of terms $\TF$ built with an alphabet ${\cal F}$ of
function symbols with arities,
a theory is given by a set ${\cal  R}$ of labeled rewrite rules
denoted $\ell(x_1,\ldots,x_n) : l \rewrite r$,
where labels $\ell (x_1,\ldots,x_n)$ record the set of variables
occurring in the rewrite rule.
Formulas are sequents of the form $\pi : \Ecl{t} \ra \Ecl{t'}$,
where $\pi$ is a {\em proof term}
recording the proof of the sequent:
${\cal R} ~\vdash~ \pi : \Ecl{t} \ra \Ecl{t'}$
if $\pi : \Ecl{t} \ra \Ecl{t'}$ can be obtained by finite application
of equational deduction rules given below.
In this context, a proof term $\pi$ encodes a
sequence of rewriting steps called a derivation.

\begin{description}
\item[Reflexivity] For any $t\in\TF$:
  {\[
  {\bf t}:\Ecl{t}\ra \Ecl{t}
  \]}
\item[Congruence] For any $f\in {\cal F}$ with  $arity(f)=n$:
  {\[
  \frac{{\bf \pi_1} : \Ecl{t_1}\ra \Ecl{t'_1}\;\;\;\ldots\;\;\;
    {\bf \pi_n} : \Ecl{t_n}\ra \Ecl{t'_n}}
       { {\bf f(\pi_1,\ldots,\pi_n)} : \Ecl{f(t_1,\ldots,t_n)} \ra \Ecl{f(t'_1,\ldots,t'_n)}}
  \]}
\item[Transitivity]
 {\[
  \frac{{\bf \pi_1}:\Ecl{t_1}\ra
    \Ecl{t_2}\;\;\;\;\;\;{\bf \pi_2}:\Ecl{t_2}\ra \Ecl{t_3}}
        {{\bf \pi_1;\pi_2}\;:\;\Ecl{t_1}\ra \Ecl{t_3}}
  \]}
 \item[Replacement] For any $\ell (x_1,\ldots,x_n) : l \rewrite r \in  {\cal R}$,
   {\[
   \frac{{\bf \pi_1} :
     \Ecl{t_1}\ra\Ecl{t'_1}\;\;\;\ldots\;\;\;{\bf \pi_n}:\Ecl{t_n}\ra\Ecl{t'_n}}
       {{\bf \ell(\pi_1,\ldots,\pi_n)} :  \Ecl{l(t_1,\ldots,t_n)} \ra \Ecl{r(t'_1,\ldots,t'_n)}}
  \]}
\end{description}

The \elan\ language, designed in 1997, introduced the concept
of strategy by giving explicit constructs for expressing control on
the rule application~\cite{borovansky02a}.
Beyond labeled rules and concatenation denoted ``$;$'', other constructs for
deterministic or non-deterministic choice, failure, iteration, were
also defined in \elan.
A strategy is there defined as a set of proof terms in rewriting logic
and can be seen as a higher-order function : 
if the strategy $\zeta$ is a set of proof terms $\pi$,
applying $\zeta$ to the term $t$ means finding all terms $t'$ such
that $\pi:\Ecl{t} \ra \Ecl{t'}$ with $\pi \in \zeta$.
Since rewriting logic is reflective, strategy semantics can be
defined inside the rewriting logic by rewrite rules at the meta-level. 
This is the approach followed by \maude\ in~\cite{Maude07,Marti-OlietN:WRLA08}.

\section{Rewriting Calculus}
\label{RewritingCalculus}

The rewriting calculus, also called $\rho$-calculus, has been introduced in 1998 by Horatiu Cirstea and Claude Kirchner~\cite{rhoCalIGLP-I+II-2001}.
As claimed on \url{http://rho.loria.fr/index.html}:

{\em The rho-calculus has been introduced as a general means to
  uniformly integrate rewriting and $\lambda$-calculus. This calculus
  makes explicit and first-class all of its components: matching
  (possibly modulo given theories), abstraction, application and
  substitutions.

The rho-calculus is designed and used for logical and semantical
purposes. It could be used with powerful type systems and for
expressing the semantics of rule based as well as object oriented
paradigms. It allows one to naturally express exceptions and
imperative features as well as expressing elaborated rewriting
strategies. }

Some features of the rewriting calculus are worth emphasizing here:
first-order terms and $\lambda$-terms are $\rho$-terms 
($\lambda x.t$ is $(x \rewrite t)$);
a rule is a $\rho$-term as well as a strategy, so
rules and strategies are abstractions of the same nature and
``first-class concepts''; application generalizes $\beta-$reduction;
composition of strategies is like function composition; 
recursion is expressed as in $\lambda$ calculus with a recursion
operator $\mu$.

In order to illustrate the use of $\rho$-calculus, let us consider 
the Abstract Biochemical Calculus (or $\rho_{Bio}$-calculus)~\cite{AndreiK-Termgraph09}.
This rewriting calculus models autonomous systems
as {\em biochemical programs} which consist of the following components:
collections of molecules (objects and rewrite rules), higher-order
rewrite rules over molecules 
(that may introduce new rewrite rules in the behaviour of the system) and
strategies for modeling the system's evolution.
A visual representation via  {\em port graphs} and an implementation are provided by the PORGY
environment described in~\cite{AndreiO:PORGY}.
In this calculus, strategies are abstract molecules, expressed with an
arrow constructor ($\Ra$ for rule abstraction), an application
operator $\appop{}{}$ and a constant operator $\stk$ for explicit failure.

Below are examples of  useful strategies in $\rho_{Bio}$-calculus:
$$
\begin{array}{rcl}
\id & \triangleq & X\Ra X \\
\fail & \triangleq & X\Ra \stk \\
\seq(S_1,S_2) & \triangleq &   X\Ra {S_2} \appop (S_1 \appop X) \\ 
\first(S_1,S_2) & \triangleq &  X\Ra ({S_1} \appop {X})\ \
{(\stk\Ra ({S_2} \appop {X}))} \appop  {(S_1 \appop X)} \\
\try(S) & \triangleq & \first(S,\id) \\
\nots(S) & \triangleq & X\Ra \first(\stk\Ra X, X'\Ra\stk) \appop ({S}\appop {X}) \\ 
\ifs(S_1,S_2,S_3) & \triangleq &  X\Ra {\first(\stk\Ra
  S_3\appop X,
  X'\Ra S_2 \appop X)} \appop {(S_1\appop {X})}\\
\repeats(S) & \triangleq &\mu X. \try(\seq(S,X)) 
\end{array}
$$

Based on such constructions, the  $\rho_{Bio}$-calculus
allows failure handling, repair instructions, persistent application
of rules or strategies,
and more generally strategies for autonomic computing,  as described
in~\cite{AndreiK08c}. 
In~\cite{AndreiK-Termgraph09}, it is shown how to do invariant
verification in biochemical programs.
Thanks to $\rho_{Bio}$-calculus, an invariant property can in many cases, be encoded
as a special rule in the biochemical program modeling the system and
this rule is dynamically checked at each execution step. 
For instance, an invariant of the system is encoded by a rule  $G\Ra G$
and the strategy verifying such an invariant is encoded 
with a persistent strategy $\first(G\Ra G, X\Ra {\stk})$.
In a similar way, an unwanted occurrence of a concrete molecule $G$
in the system can be modeled with the rule
$(G\Ra {\sf stk})$. 
And instead of yielding failure ${\stk}$, the problem
can be ``repaired'' by associating to each property the necessary
rules or strategies to be inserted in the system in case of failure.

\section{Abstract Reduction Systems}
\label{ARS}

Another view of rewriting is to consider it as an abstract relation on
structural objects.
An \emph{Abstract Reduction System (ARS)}~\cite{TereseStrategies2003,KKK08,BCDK-WRS09} is a labeled oriented graph $(\cal O,
  \cal S)$ with a set of labels ${\cal L}$. The nodes in $\cal O$ are called \emph{objects}.
   The oriented labeled edges in $\cal S$  are called \emph{steps}: 
$a \flechup{\phi}  b$ or $(a,\phi,b)$, with \emph{source} $a$, \emph{target} $b$ and \emph{label} $\phi$.
Derivations are composition of steps.  

For a given ARS $\cal A$, an \emph{$\cal A$-derivation} is denoted
$\pi : a_0 \flechup{\phi_0} a_1
 \flechup{\phi_1} a_2 \ldots \flechup{\phi_{n-1}} a_n$
 or $a_0 \flechup{\pi} a_n$, where $n \in \Nat$.
The \emph{source} of $\pi$ is $a_0$ 
and its domain $Dom({\pi})=\{a_0\}$.
The \emph{target} of $\pi$ is $a_n$
and applying $\pi$ to $a_0$ gives the singleton set $\{a_n\}$,
which is denoted $\pare{\pi} a_0=\{a_n\}$.

Abstract strategies are defined in~\cite{KKK08} and
in~\cite{BCDK-WRS09} as follows:
for a given ARS $\cal A$, an \emph{abstract strategy} $\zeta$ is a subset of the set of
    all derivations (finite or not) of $\cal A$.
The notions of domain and application are generalized as follows:
$Dom(\zeta) = \bigcup_{\pi \in \zeta} Dom(\pi)$
and
$\pare{\zeta} a
    = \{b ~|~ \exists \pi \in \zeta \mbox{ such that } a \flechup{\pi}
    b\}  = \{\pare{\pi} a ~|~ \pi \in \zeta\}$. 
Playing with these definitions,~\cite{BCDK-WRS09} explored adequate definitions of
termination, normal form and confluence under strategy.

Since abstract reduction systems may involve infinite sets of objects, of reduction steps
and of derivations, we can schematize them with constraints at different levels: (i) to describe the objects occurring in a derivation (ii) to describe, via the
labels, requirements on the steps of reductions (iii) to describe the structure of the
derivation itself (iv) to express requirements on the histories.
The framework developed in~\cite{KKK-wfpl09} defines a strategy $\zeta$ as all instances
$\sigma(D)$ of a derivation schema $D$ such that $\sigma$ is solution of a constraint
$C$ involving derivation variables, object
variables and label variables. As a simple example, the infinite set of derivations of length one that
transform  $a$ into $f(a^n)$ for all $n \in \Nat$, where $a^n=a*\ldots*a$ ($n$
times), is simply described by:
$(a \ra f(X) \mid X * a =_{A} a * X) $, where $=_{A}$
indicates that the constraint is solved modulo associativity of
the  operator $*$.
This very general definition of abstract strategies is called {\em  extensional} in~\cite{BCDK-WRS09}
in the sense that a strategy is defined explicitly as a set of derivations of an abstract
  reduction system.  The concept is  useful to
  understand and unify reduction systems and deduction systems as
  explored in~\cite{KKK08}.

But abstract strategies do not capture another point of view,
also frequently adopted in rewriting:
a strategy is a partial function that associates to a
reduction-in-progress, the possible next steps in the reduction
sequence.  
Here, the strategy as a function depends only on the object and the
derivation so far.  This notion of strategy coincides with the
definition of strategy in sequential path-building games,
with applications to planning, verification and
synthesis of concurrent systems~\cite{Dougherty08}.
This remark leads to the following  \emph{intensional} definition given in~\cite{BCDK-WRS09}.
The essence of the idea is that strategies
are considered as a way of constraining and guiding the steps of a
reduction. So at any step in a derivation, it should be
possible to say whether a contemplated next step obeys the strategy $\zeta$.
In order to take into
account the past derivation steps to decide the next possible ones,
the history of a derivation has to be memorized and available at
each step. Through the notion of traced-object
$\cx{\alpha}a = \cx{(a_0,\phi_0),\ldots,(a_n,\phi_n)} a$ in $\cobj$, each object $a$
memorizes how it has been reached with the trace $\alpha$.

An \emph{intensional strategy}  for ${\cal A = (O,S)}$ is a partial
function $\nzeta$ from $\cobj$ to $2^{\cal S}$ such that 
 for every traced object $\cx{\alpha}a$, $\nzeta(\cx{\alpha}a) \subseteq
  \{\pi\in {\cal S} \mid Dom(\pi)=a \}$.
If $\nzeta(\cx{\alpha} a)$ is a singleton, then the reduction step
under $\nzeta$ is deterministic.

As described in~\cite{BCDK-WRS09}, an intensional strategy $\nzeta$ naturally
generates an abstract strategy, called its \emph{extension}: this is the abstract
  strategy $\ezeta$ consisting of the following set of derivations:\\
$\forall n \in \Nat$, $\pi : a_0 \flechup{\phi_0} a_1
 \flechup{\phi_1} a_2 \ldots \flechup{\phi_{n-1}} a_n  \; \in \ezeta
\qquad \text{ iff } \qquad \forall j\in [0,n], \quad
(a_j \flechup{\phi_{j}} a_{j+1}) \in \nzeta(\cx{\alpha}a_{j})
$.\\
This  extension may obviously contain infinite derivations; in
such a case it also contains all the finite derivations that are
prefixes of the infinite ones, and so is closed under taking prefixes. 

A special case are memoryless strategies, where the function $\nzeta$ 
does not depend on the history of the objects. This is the case of 
many strategies used in rewriting systems, as shown in the next example.
Let us consider an \ars\ ${\cal A}$ where objects are terms, reduction is term rewriting 
with a rewrite rule in the rewrite system,
and labels are positions where the rewrite rules are applied.
Let us consider an order $<$ on the labels which is the prefix order on positions.
Then the intensional strategy that corresponds to innermost rewriting is 
    $\nzeta_{inn}(t) = \{\pi: t \flechup{p} t' \mid p=max(\{p'
    \mid t \flechup{p'} t' \in {\cal S} \})\}$.
When a lexicographic order is
 used, the classical \emph{rightmost-innermost} strategy is obtained.

Another example, to illustrate the interest of traced objects, is the intensional strategy that restricts the derivations to be of
    bounded length $k$. Its definition makes use of the size of the trace $\alpha$,
    denoted $|\alpha|$:
    $\nzeta_{ltk}(\cx{\alpha}a) =  \{\pi \mid \pi\in {\cal S}, \;
    Dom(\pi)=a,  \;|\alpha| < k-1\}$.
However, as noticed in~\cite{BCDK-WRS09}, the fact that intensional strategies 
generate only prefix closed
abstract strategies prevents us from computing abstract strategies
that look straightforward: there is no intensional strategy that can generate a set of derivations of
length exactly $k$. Other solutions are provided in~\cite{BCDK-WRS09}.

\section{Conclusion}

A lot of interesting questions about strategies are yet open, going from the
definition of this concept and the interesting properties we may expect
to prove, up to the definition of domain specific strategy languages.
As further research topics, two directions seem really interesting to
explore:\\
- The connection with Game theory strategies.
In the fields of system design and verification,
 \emph{games} 
 have emerged as a key tool.  Such games have been studied since the
 first half of 20th century in descriptive set theory \cite{Kechris95},
 and 
 they have been adapted and generalized for applications in formal
 verification; introductions can be found in
 \cite{dagstuhl2001,Walukiewicz04}.  
It is worth wondering whether the coincidence of the term ``strategy'' in the domains
of rewriting and games is more than a pun.  It should
be fruitful to explore the connection and to be
guided in the study of the foundations of strategies by some of the
insights in the literature of games. \\
- Proving properties of strategies and strategic reductions. A lot of
work has already begun in the rewriting community and have been 
presented in journals, workshops or conferences of this domain.
For instance, properties of confluence, termination, or completeness
for rewriting under strategies have been addressed, either
based on schematization of derivation trees, as
in~\cite{GK-TOCL09},
or by tuning proof methods to handle specific strategies  (innermost, outermost, lazy
  strategies) as in~\cite{Giesl-Mid-inn-context-sens-2003, Giesl2011}.
Other approaches as~\cite{Balland2012} use strategies
transformation to equivalent rewrite systems to be able to reuse
well-known methods.
Finally, properties of strategies such as fairness or  loop-freeness
could be worthfully explored by making connections between 
different communities (functional programming, proof theory,
verification, game theory,...).

\paragraph{Acknowledgements} The results presented here are based on
pioneer work in the \elan\ language designed in the Protheo team
from 1997 to 2002.  They rely on joint
work with many people, in particular Marian Vittek and Peter Borovansk{\'y}, Claude Kirchner and Florent
Kirchner, Dan Dougherty, Horatiu Cirstea and Tony Bourdier, Oana Andrei, Maribel
Fernandez and Olivier Namet.
I am grateful to Jos\'e Meseguer and to the members of the PROTHEO
and the PORGY teams, for many inspiring discussions on the topics of
this talk.

\bibliographystyle{eptcs} 
\bibliography{biblio}

\end{document}

%% file: minmacros.tex
\newcommand{\nzeta} {\ensuremath{\lambda}}

\newcommand{\ezeta} {\ensuremath{\zeta_{\nzeta}}}

\newcommand{\TF}[0]	{{\cal T(F)}}

\newcommand{\ars}{abstract reduction system}

\newcommand{\rew}[3]{\mathop{{\stackrel{\small #1}{\longrightarrow}}_{#3}^{#2}}}

\newcommand{\ra}{\rightarrow}

\newcommand{\flechup}[1]{\xrightarrow{#1}}

\newcommand{\Ra}{\Rightarrow}
\newcommand{\rewrite}{\Rightarrow}
\newcommand{\Ecl}[1]{#1}

\newcommand{\appop}{{{\scriptscriptstyle {}^{\bullet}}}} 

\newcommand{\cx}[1]{\ensuremath{\left[#1\right]\vspace{1pt}}}
\newcommand{\pare}[1]{\ensuremath{#1 \vspace{1pt} \appop \vspace{1pt}}}

\newcommand{\Ee}{{\cal O}}

\newcommand{\cobj}{\ensuremath{\Ee^{[\ca A]}}}

\newcommand{\ca}[1]{\ensuremath{\mathcal{#1}}}

\newcommand{\elan}    {\textsf{ELAN}}
\newcommand{\maude}   {\textsf{Maude}}

\newcommand{\Nat}{\mathbb N}

\newcommand{\stk}{{\sf stk}}

\newcommand{\id}{{\sf id}}

\newcommand{\ifs}{{\tt ifTE}}
\newcommand{\seq}{{\tt seq}}
\newcommand{\fail}{{\tt fail}}
\newcommand{\nots}{{\tt not}}

\newcommand{\first}{{\tt first}}

\newcommand{\try}{{\tt try}}
\newcommand{\repeats}{{\tt repeat}}

